\documentclass[12pt]{article}

\usepackage[utf8]{inputenc}
\usepackage[T1]{fontenc}
\usepackage{newtxtext} 

\usepackage{amsmath, amsfonts, amsthm, amsgen}
\usepackage{amssymb}   
\usepackage{newtxmath} 

\usepackage{bm}        
\usepackage{mathrsfs}  
\usepackage{dsfont}    
\usepackage{calc}

\usepackage[margin=1in]{geometry}
\usepackage{setspace}

\usepackage{xcolor}
\usepackage{graphicx}
\usepackage{subcaption}
\usepackage{rotating}

\usepackage{tikz}
\usetikzlibrary{arrows.meta, positioning}

\usepackage{pstricks, multido, pst-node}

\usepackage{array}
\usepackage{booktabs}
\usepackage{multirow}
\usepackage{longtable}

\usepackage[numbers]{natbib}
\setcitestyle{citesep={,}}

\usepackage{url}
\usepackage[colorlinks,citecolor=blue,urlcolor=blue]{hyperref}

\newcommand{\ci}{\perp\!\!\!\perp}

\newenvironment{keyword}%
  {\vspace{1em}\noindent\textbf{Keywords:}\quad}%
  {\par}

\newcommand{\kwd}[1]{#1; }

\theoremstyle{plain}
\newtheorem{assumption}{Assumption}[section]

\usepackage{fancyhdr}
\begin{document}

\begin{center}
    {\Large \bfseries A fully Bayesian causal factor model to evaluate England's Hepatitis C peer support program} \\[0.3in]

    {\large
    Constantin Schmidt (MSc)$^{1\dagger}$,
    Pantelis Samartsidis (PhD)$^{1}$,
    Shaun R.\ Seaman (PhD)$^{1}$,
    Beatrice Emmanouil (PhD)$^{2,3}$,
    Graham R.\ Foster (PhD)$^{3}$,
    Leila Reid (MSc)$^{4}$,
    Stuart Smith (MSc)$^{4}$,
    Daniela De Angelis (PhD)$^{1}$
    } \\[0.2in]

    \textit{
    $^1$MRC Biostatistics Unit, University of Cambridge, East Forvie Building, Robinson Way, Cambridge CB2 0SR; \\
    $^2$National Health Service England, Wellington House, 133-155 Waterloo Road, London, SE1 8UG; \\
    $^3$Blizard Institute, Faculty of Medicine and Dentistry, Queen Mary University of London, 4 Newark Street, London, E1 2AT; \\
    $^4$Hepatitis C Trust, 72 Weston St, London SE1 3QH.
    } \\[0.2in]

    {$^\dagger$Corresponding author: \href{mailto:constantin.schmidt@mrc-bsu.cam.ac.uk}{constantin.schmidt@mrc-bsu.cam.ac.uk}; +44 7897 269482.}
\end{center}

\vspace{0.1in}

\begin{abstract}
Starting in 2018, National Health Service England employed peer support workers (peers) to boost Hepatitis C virus (HCV) elimination efforts. Peers are individuals with relevant lived experience who educate their communities and promote testing and treatment. We assess the causal effect of the peers intervention on case-finding of HCV-infected individuals using a new, fully Bayesian causal factor analysis model.
Our method refines existing approaches in several ways: First, it improves coverage of credible intervals of causal estimands by jointly modelling the intervention assignment process, pre- and post-intervention outcomes. Second, it provides estimates of both conditional average and individual treatment effects (ITEs). For ITEs, we propose a copula-based approach that allows to account for uncertainty about assumptions made regarding the joint distribution of potential outcomes. Third, our model is applicable to ordinal staggered interventions and count-valued outcome data.
Our analysis suggests that the introduction of peers led to an increase in case-finding of HCV-infected individuals. Further, we found that the effect of the intervention increased with intervention intensity, and was stronger during the national COVID-19 lockdown.
\end{abstract}

\vspace{0.1in}

\textbf{Acknowledgments}
CS is funded by the Medical Research Council (MRC) Biostatistics Unit Core Studentship.
SRS is funded by MRC grant MC\_UU\_00040/05.
DDA is funded by MRC grant MC\_UU\_00002/11.
PS is funded by MRC grant UKRI332.


\textbf{Conflict of Interest}
GF has previously received funding from companies that market  antiviral drugs for hepatitis C --- Abbvie, Gilead, MSD.



\newpage
\doublespacing

\section{Introduction}\label{sec:intro}

Hepatitis C virus (HCV) is a blood-borne virus that affects the liver. 
When left untreated, HCV infection can lead to to acute liver damage, cirrhosis, cancer, and eventually death \cite{uk_health_security_agency_hepatitis_2024}.
HCV is a major public health concern worldwide. 
In 2022, the World Health Organization (WHO) estimated that approximately 50 million people globally were living with an HCV infection \cite{world_health_organisation_hepatitis_2024}. 
In England, the burden caused by HCV is also significant, with an estimated 62,600 adults living with a chronic HCV infection as of 2022 \cite{uk_health_security_agency_hepatitis_2024}.

Since 2015, the introduction of highly effective -- over 95\% cure rate -- and well-tolerated direct-acting antiviral (DAA) drugs has facilitated international efforts to eliminate HCV \cite{asselah_treatment_2018}. 
Several WHO signatories, including the United Kingdom, have committed to an elimination strategy aiming to reduce new HCV infections by 90\% and HCV-related mortality by 65\% by 2030, compared to 2016 levels \cite{world_health_organisation_elimination_2025}.

To achieve these targets, DAA treatment must be made accessible to everyone, especially those who are at high risk of HCV infection.
However, most HCV infections in high income countries occur in people who are poorly engaged by traditional health services.
One such example is people who inject drugs (PWID). 
Worldwide, approximately 5.8 million PWID are infected with HCV \cite{world_health_organisation_hepatitis_2024}. 
In England, over 80\% of HCV-infected individuals are current injecting drug users or people with a history of injecting drug use \cite{uk_health_security_agency_hepatitis_2025}. 
Other high-risk populations include individuals with a history of incarceration, those experiencing homelessness, or people who grew up in a country with a high prevalence of HCV infection.

Several `elimination initiatives’ have been introduced in England to improve anti-HCV treatment coverage among these high risk populations. 
One such initiative is the provision of peer support workers, or \textit{peers}. 
Peers are individuals from the community with relevant lived experience who educate their communities about HCV and encourage testing and treatment. 
In January 2018, the \textit{Hepatitis C Trust} (HCT), a London-based charity dedicated to HCV, began employing peers to work with HCV healthcare teams across England.
HCV healthcare teams in England are organised into 22 regional networks called Operational Delivery Networks (ODNs).
A rigorous evaluation of the peers intervention has not yet been carried out.
In this study, we address this gap in the literature by assessing the causal effect of the peers intervention on case-finding of HCV-infected individuals, using data up to May 2021. 

This evaluation poses multiple challenges.
First, this is a non-randomised intervention, implying that there is potential for confounding.
Second, the intervention started at different time points across the ODNs (staggered adoption), which necessitates adjustment for temporal variations in treatment uptake and the outcome. 
Third, the number of peers operating at each ODN and time point varied, resulting in different intervention intensities and potentially heterogeneous intervention effects. 
Fourth, our outcome of interest is an over-dispersed count outcome, which introduces modelling challenges.
Fifth, there is only a small number of ODNs, which makes defining and estimating a meaningful estimand challenging, as discussed in more detail below.

Non-randomised interventions with staggered adoption are typically evaluated using counterfactual imputation models \cite{liu_practical_2022}.
These methods `impute' the outcomes that would have been observed in the post-intervention time periods had the intervention not occurred.
To do this, they build a counterfactual prediction model using data from units -- in our application ODNs -- that are not exposed to the intervention and pre-intervention data from eventually exposed units.

Counterfactual imputation approaches include synthetic control \cite{abadie_synthetic_2010, brodersen_inferring_2015, xu_generalized_2017}, and causal factor analysis models (or `causal matrix completion'; \cite{athey_matrix_2021, pang_bayesian_2022, nethery_integrated_2023, samartsidis_bayesian_2024}), which account for the potential of both observed and unobserved confounding in different ways.
For an overview of these methods and the assumptions they make see for example Samartsidis et al. \cite{samartsidis_assessing_2019} or Liu, Wang, and Xu \cite{liu_practical_2022}. 
One disadvantage of counterfactual imputation models is that they discard part of the available information -- namely the post-intervention data of eventually exposed units and intervention assignment -- from parameter estimation. 
In addition, existing methods are either not applicable to count data or unable to deal with non-binary interventions, both of which are essential features of our application.  

Counterfactual imputation models generally provide estimates of the individual treatment effects (ITEs) or estimates of treatment effects that summarise several ITEs.
ITEs quantify the impact that the intervention had at each time point during an exposed unit's post-intervention period.
ITEs are obtained as the difference between the observed post-intervention outcomes and imputed intervention-free outcomes.
ITEs have a straightforward interpretation, which can be valuable to policy makers.
However, estimating ITEs requires modelling the joint distribution of potential outcomes -- intervention-free and under intervention -- including modelling the associations between potential outcomes for the same unit at the same time period \cite{ding_causal_2018}.
Since only one potential outcome from each unit can ever be observed for one specific time period, the associations between potential outcomes cannot be estimated from the data.
Thus, several authors have suggested to vary the parameter governing the associations between potential outcomes in a plausible range as a sensitivity analysis \cite{richardson_transparent_2011, ding_potential_2016}.
However, existing counterfactual imputation approaches assume potential outcomes are independent conditional on some covariates for modelling convenience without providing any means of investigating the sensitivity of results to this fundamental assumption.

An alternative to ITEs are conditional average treatment effects (CATEs) quantifying the average intervention effect conditional on effect modifiers and characteristics of the intervention in a population of units \cite{li_bayesian_2023}.
Since we are dealing with count data, we are specifically interested in conditional average risk ratios. 
The estimation of CATEs does not require assumptions regarding the joint distribution of potential outcomes.
However, in applications in which the population of units is small and fully observed, for example, in our application the sample of ODNs is exhaustive, CATEs are challenging to interpret.
CATEs refer to a super-population of units, which, if all units are observed, is hard to conceptualise.
CATEs are popular in the analysis of stepped-wedge cluster randomised trials, which are staggered intervention designs with randomised intervention start times \cite{kenny_analysis_2022}.
The literature on estimating CATEs from non-randomised interventions with staggered adoption is limited. 

The contributions of this paper are twofold. 
The first contribution is to propose a principled Bayesian methodology to evaluate non-randomised interventions with staggered adoption based on causal factor analysis.
Our method addresses three challenges of current state-of-the-art methods discussed above.
First, our model makes use of all available data (pre- and post-intervention outcomes and intervention assignment) to inform parameter estimation.
We show in a simulation study that not modelling intervention assignment may lead to poor coverage of credible intervals.
Second, our model provides estimates of both ITEs and CATEs. 
For ITEs, we propose a copula-based approach that allows the practitioner to explore sensitivity of the results to the value specified for the correlation parameter between potential outcomes, or to account for uncertainty in this parameter.
We provide a detailed R script for this approach.
Third, our model is applicable to ordinal staggered interventions and count-valued outcome data, which is essential for our application. 
We develop our method under the Bayesian paradigm, which allows full characterisation of uncertainty for all the causal estimands of interest.

The second contribution is the application of the proposed methodology to evaluate the effect of the peers intervention on case-finding of HCV-infected individuals in England. 
Our analyses provide invaluable insights on the effectiveness of this intervention, which we believe have implications for similar interventions in the future. 
We compare the results obtained through our method with results obtained through a state-of-the art counterfactual imputation factor analysis model \cite{nethery_integrated_2023, samartsidis_bayesian_2024}.

The remainder of this paper is organised as follows: 
In Section \ref{sec:Application}, we provide some background on the intervention, describe our data sources and list the epidemiological questions of interest. 
In Section \ref{sec:BMCM}, we develop a methodology that can be used to evaluate the peers intervention. 
In Section \ref{sec:Results}, we presents the evalution results. 
Finally, in Section \ref{sec:Discussion3} we conclude with a discussion and list some possible directions for future research.
The Supplementary Material contains additional implementation details, results, and a simulation study.
Replication code is available at \url{https://github.com/constantin-schmidt/HCV_peers.git} including an R script showing how to implement our copula based approach to account for uncertainty regarding the joint distribution of potential outcomes for a generic model.

\section{Background}
\label{sec:Application}
The peers intervention is one of the initiative introduced by National Health Service England (NHSE) as part of its HCV elimination programme. 
Its implementation was motivated by earlier research demonstrating the value of peer support in other areas of disease such as the human immunodeficiency virus \cite{magidson_someone_2019}, and in improving the engagement of HCV-infected individuals with treatment in two small controlled trials \cite{stagg_improving_2019,ward_randomized_2019}. 

In 2018, each of the ODNs was offered the opportunity to appoint a peer fully funded by NHSE. 
Prior to initiating their job, the peers received specialised training from the HCT. 
To prevent excess requests for training, it was decided that peers would be introduced to ODNs gradually. 
The order in which peers were appointed in ODNs was not randomised; instead it was determined by factors associated with the ODNs' readiness and willingness to introduce the intervention. 
Following this initial phase of recruitment (roughly two years), additional peers were appointed in ODNs depending on availability of funds and operational readiness. 

Peers work closely with clinical teams and are involved in various stages of HCV prevention, diagnosis and treatment. 
An important aspect of their job is to perform outreach, particularly in services working with marginalised individuals at high risk of HCV infection, such as drug services, needle exchange centres and homeless hostels. 
There, they deliver workshops to educate local communities about HCV, offer HCV testing and link HCV-infected individuals to the clinical teams to facilitate treatment. 
Peers also support individuals for whom clinical teams might otherwise struggle to ensure engagement -- for example due to ongoing drug use -- throughout the treatment process (roughly twelve weeks). 
Finally, peers are responsible for recruiting volunteers to assist them in these activities.

The objective of this paper is to evaluate the impact of the peers intervention on the number of DAA-eligible individuals identified (henceforth \textit{case-finding} for brevity).
We measure case-finding by the total number of DAA therapy funding requests submitted to the NHSE system.
These funding request are mandatory for each individual with confirmed HCV infection which a clinical team has deemed eligible for DAA therapy.
We chose this outcome due to the availability of well-recorded data and its role as an important first step in the treatment process.

We are interested in answering the following three questions regarding the impact of peers on case-finding. 
First, whether the introduction of peers led to an increase in the number DAA-eligible individuals identified. 
Second, whether the effect (if any) depended on intervention intensity.
Here intensity refers to the cumulative number of peer-months since the intervention's introduction.
We expect that during the initial stages of the intervention, some effort is required by peers to setup their outreach plans before they are able to develop connections with HCV-infected individuals. 
Third, whether the effect (if any) of the intervention was stronger during the first COVID-19 national lockdown in England (March-May 2020). 
This question is motivated by anecdotal evidence highlighting the extended role of peers during the lockdown, when access to healthcare was limited.

\begin{figure}[ht]
    \centering
    \begin{minipage}[b]{0.49\linewidth}
        \centering
        \includegraphics[width=6.5cm]{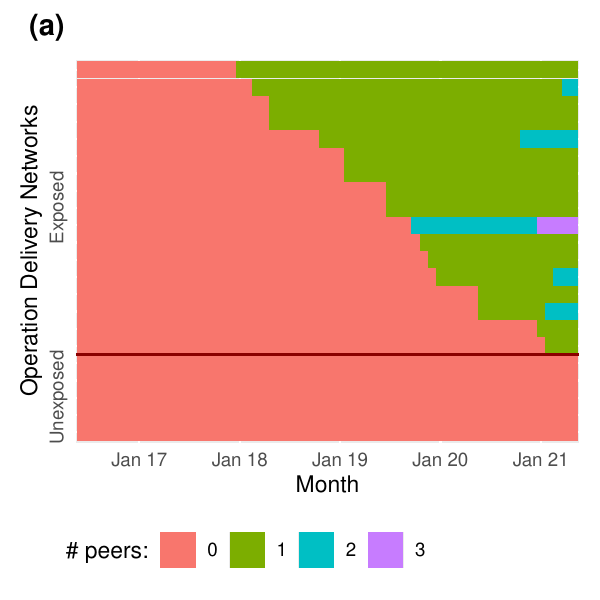}
        \par\vspace{1ex}
        (a)
    \end{minipage}
    \hfill
    \begin{minipage}[b]{0.49\linewidth}
        \centering
        \includegraphics[width=6.5cm]{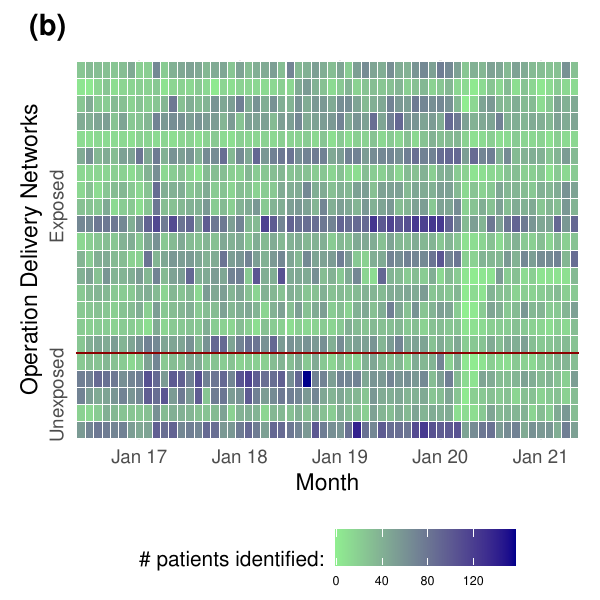}
        \par\vspace{1ex}
        (b)
    \end{minipage}
    \caption{Graphical summaries of data. Abbreviations: Jan: January; \#: Number. 
    (a) Number of peers working in each Operation Delivery Network during each month between June 2016 and May 2021.
    (b) Number of individuals living with Hepatitis C infection and eligible for treatment with direct-acting antiviral drugs identified each month between June 2016 and May 2021.}
    \label{fig:DataSummary}
\end{figure}

We use monthly data covering the period from January 2016 to May 2021. 
Information on the total number of paid peers working at each ODN over time was provided by the Hepatitis C Trust, which is responsible for peer recruitment and training. 
Figure \ref{fig:DataSummary}(a) displays the total number of paid peers at each ODN over time. 
The first paid peer began working in January 2018.
By the end of our study period, five ODNs had not employed a peer, while five others had employed more than one.
Data on the case-finding in each ODN over time were obtained from Bluteq, an online portal used by NHSE to manage high-cost medicine funding requests \cite{blueteq_ltd_blueteq_2025}, see Figure \ref{fig:DataSummary}(b). 
Over the study period, there were a total of 56,137 DAA-eligible individuals identified, corresponding to an average of 42.53 HCV-infected individuals per ODN per month. 
Notably, case-finding dropped sharply across all ODNs during the first COVID-19 national lockdown in England.

\section{A Bayesian causal factor analysis model for ordinal staggered interventions}
\label{sec:BMCM}

In this section, we develop the approach that we will later use to assess the peers intervention described in Section \ref{sec:Application}: we
 introduce the causal framework and causal estimands of interest (Section \ref{subsec:causal}); we 
 list the main assumptions that enable identification of these estimands from the data (Section \ref{subsec:BFM_modelassumptions}); we outline estimation under the Bayesian paradigm (Section \ref{subsec:bayes}); and 
 we discuss some implementation details (Section \ref{subsec:computations}).

We first introduce some notation.
We use upper-case letters for random variables and lower-case letters for their possible values.
Let there be $i = 1, \ldots, N$ units and $t = 1, \dots, T$ time periods relative to a common calendar time. 
For each unit-time pair $\{i,t\}$, the observed data consists of the outcome of interest $Y_{it}$ (number of DAA-eligible individuals identified) and the intervention intensity $A_{it}$ (number of peers). 
Let $G_i$ be the first time period a unit is exposed to the intervention (the first time $A_{it}>0$).
As a convention, we set $G_i = T+1$ if $i$ is a never exposed (henceforth `control') unit.
For generality of exposition, we further introduce a vector of covariates $\boldsymbol{X}_{it}$, although these are not available in our specific application. 
For any variable $Z_{it}$, define $\bar{\boldsymbol{Z}}_{it} = (Z_{i1}, \ldots, Z_{it})^{\top}$ as the history of this variable up to time $t$. 
Let $\boldsymbol{0}_t$ be the $t$-vector with all elements zero.

\subsection{Causal framework and estimands}
\label{subsec:causal}
We use the potential outcomes framework \cite{rubin_causal_2005} for causal inference. 
In this framework, for each pair $\{i,t\}$, we define the potential outcome $Y_{it}\left( \bar{\boldsymbol{a}}_{1T}, \ldots, \bar{\boldsymbol{a}}_{nT} \right)$ as the outcome that unit $i$ would experience at time $t$ if the intervention were rolled out as indicated by $\bar{\boldsymbol{a}}_{1T}, \ldots, \bar{\boldsymbol{a}}_{nT}$. 
We now lay out some assumptions which are standard in the causal inference literature.

\begin{assumption} \label{ass:NoInterference}
    (No interference). 
    For all $i$ and $t$, $Y_{it}\left( \bar{\boldsymbol{a}}_{1T}, \ldots, \bar{\boldsymbol{a}}_{nT} \right) = Y_{it}\left( \bar{\boldsymbol{a}}_{1T}^*, \ldots, \bar{\boldsymbol{a}}_{nT}^* \right)$ for any 
    $\bar{\boldsymbol{a}}_{1T}, \ldots, \bar{\boldsymbol{a}}_{nT}$
    and $\bar{\boldsymbol{a}}_{1T}^*, \ldots, \bar{\boldsymbol{a}}_{nT}^*$
    such that $\bar{\boldsymbol{a}}_{iT} = \bar{\boldsymbol{a}}_{iT}^*$.
\end{assumption}  

Assumption \ref{ass:NoInterference} states that the potential outcomes in unit $i$ are not affected by intervention roll-out in other units. 
This allows us to simplify the notation for potential outcomes to $Y_{it}(\bar{\boldsymbol{a}}_{iT})$. 
No interference is a realistic assumption in our motivating application since peers operate locally within their ODNs and are thus unlikely to affect case-finding in other ODNs. 

\begin{assumption} \label{ass:NonIrreversibility}
    (Irreversibility of intervention). 
    For all $i$ and $t$,
        $a_{it} \geq a_{it-1}$.
\end{assumption}  

As can be seen in Figure \ref{fig:DataSummary}(a), Assumption \ref{ass:NonIrreversibility} is satisfied in our data.
Let $\bar{\mathcal{A}}_t$ be the set of all possible intervention paths up to $t$ that do not violate Assumption \ref{ass:NonIrreversibility}.

\begin{assumption} \label{ass:NEPT}
    (No anticipation). 
    For all $i$ and $t$, $Y_{it}(\bar{a}_{iT}) = Y_{it}(\bar{a}_{iT}^\ast)$ for any $\bar{a}_{iT}$ and $\bar{a}_{iT}^\ast$ such that $\bar{a}_{it} = \bar{a}_{it}^\ast$.
\end{assumption}

Assumption \ref{ass:NEPT}, states that changes in intervention intensity cannot affect the outcome prior to their introduction. 
In our motivating application, Assumption \ref{ass:NEPT} applies as ODNs did not alter their HCV treatment pipelines in anticipation of a peer. 
Invoking Assumption \ref{ass:NEPT} allows us to further simplify notation for potential outcomes to $Y_{it}(\bar{\boldsymbol{a}}_{it})$.

\begin{assumption} \label{ass:consistency}
    (Consistency). 
    For all $i$ and $t$,
        $Y_{it} = \sum_{\bar{\boldsymbol{a}}_{t} \in \bar{\mathcal{A}}_t} 
        Y_{it}(\bar{\boldsymbol{a}}_{t}) \mathds{1}\{\bar{\boldsymbol{A}}_{it}=\bar{\boldsymbol{a}}_t \}$. 
\end{assumption} 

Assumption \ref{ass:consistency} links the potential outcomes to the outcomes observed in the data.
It states that the potential outcome $Y_{it}(\bar{\boldsymbol{a}}_{t})$ that would be observed if $\bar{\bm{A}}_{it}$ was set to what it in fact was is equal to the observed value of $Y_{it}$ \cite{vanderweele_causal_2013}.

We can now express causal estimands that will allows us to address the questions listed in Section \ref{sec:Application} in terms of potential outcomes. 
For each $i$ and $t\geq G_i$, we define the ITE as
\begin{equation}\label{eq:ite}
    \tau_{it} = Y_{it}(\bar{\boldsymbol{A}}_{it})-Y_{it}(\boldsymbol{0}_t).
\end{equation}
The $\tau_{it}$ in Equation \eqref{eq:ite} quantifies the effect that following intervention path $\bar{\boldsymbol{A}}_{it}$ has on case-finding of HCV-infected individuals for ODN $i$ at time $t$, compared to the scenario in which the ODN did not introduce any peers up to time $t$. 
Since there are many possible intervention paths up to time $t$, various other contrasts between potential outcomes could be defined.
We can also define estimands that summarise the $\tau_{it}$s. 
Here, we consider: the total number of additional DAA-eligible individuals identified thanks to the peers intervention (henceforth cumulative intervention effect),
\begin{equation}\label{eq:itecumulative}
    \tau = \sum_{i,t:A_{it}>0} \tau_{it},
\end{equation}
and the total number of additional DAA-eligible individuals identified during the first national COVID-19 lockdown,
\begin{equation}\label{eq:itecovid}
   \tau_c = \sum_{i,t: A_{it}>0, t_0\leq t\leq t_1} \tau_{it},  
\end{equation}
where $t_0$ and $t_1$ are the earliest and latest time points, respectively, during which the lockdown was in effect. 
It is possible to express estimands in terms of the \% increase, by considering
\[
\chi_{it}=100 \times \frac{\tau_{it}}{Y_{it}(\boldsymbol{0}_t)}.
\]
$\chi_{it}$ quantifies the percentage increase in case-finding of HCV-infected individuals for ODN $i$ at time $t$ comparing following intervention path $\bar{\boldsymbol{A}}_{it}$ to the scenario in which the ODN did not introduce any peers up to time $t$. 
As an overall effect in terms of percentage increase, we consider:
\[
\chi = 100 \times 
\frac{ \sum_{i,t:A_{it}>0} \tau_{it}}{\sum_{i,t:A_{it}>0} Y_{it}(\boldsymbol{0}_t)},
\]
$\chi$ quantifies the percentage increase in case-finding of HCV-infected individuals due to the peers intervention across all intervention ODNs for all post-intervention time periods.
Such effects can be easier to communicate as they do not require an understanding of what constitutes a high/low number of HCV-infected individuals identified.

As population average treatment effects, we consider the rate ratios
\begin{equation}\label{eq:psi}
\omega_t(\bar{\boldsymbol{a}}_t) = \frac{E\left[Y_{it}(\bar{\boldsymbol{a}}_t)\right]}{E\left[Y_{it}(\boldsymbol{0}_t)\right]}.
\end{equation}
This can be interpreted as the ratio between expected case-finding under the intervention sequence $\bar{\boldsymbol{a}}_t$ and expected case-finding under the reference sequence $\boldsymbol{0}_t$ (no peers), in the population of units from which our sample has been drawn. 
We can use $\omega_t(\bar{\boldsymbol{a}}_t)$ to investigate the role of cumulative exposure (by choosing appropriate $\bar{\boldsymbol{a}}_t$), and the effectiveness of the peers intervention during the lockdown (by choosing $t_0\leq t \leq t_1$). 
Due to the large number of possible intervention paths, summarising the $\omega_t(\bar{\boldsymbol{a}}_t)$s is challenging.

When there are covariates, we are further interested in calculating the rate ratios of potential outcomes conditional on covariates, which are analogous to CATEs. 
These conditional rate ratios are denoted as $\omega_t(\bar{\boldsymbol{a}}_t,\boldsymbol{x}_{it})$ and defined as:
\[
\omega_t(\bar{\boldsymbol{a}}_t,\boldsymbol{x}_{it})=\frac{E\left[Y_{it}(\bar{\boldsymbol{a}}_t)\mid\boldsymbol{x}_{it}\right]}{E\left[Y_{it}(\boldsymbol{0}_t)\mid\boldsymbol{x}_{it}\right]}.
\]
 
\subsection{Identifying assumptions}
\label{subsec:BFM_modelassumptions}

Due to the non-randomised nature of the peers intervention, it is necessary to adjust for potential confounding when estimating the causal effects defined in Section \ref{subsec:causal}. 
To carry out these adjustments, one must make assumptions regarding the relationship between intervention assignment mechanism and potential outcomes. 
A standard assumption in cross-sectional studies is that the assignment of the intervention and the potential outcomes are independent conditional on the observed covariates (selection on observables, \cite{li_bayesian_2023}). 
This is not a realistic assumption in our application due to the existence of unmeasured variables that we know were taken into account by ODNs prior to introducing a peer, and likely affect case-finding.
These include, but are possibly not limited to, the total number of HCV-infected individuals within each ODN, behavioural characteristics of HCV-infected individuals (e.g.\ risk behaviour) and the perceived administrative capacity of an ODN to integrate a peer. 
We will denote these unmeasured variables by $\boldsymbol{U}_{it}$.  

Specifically, we assume selection on observed covariates $\bm{X}_{it}$ and unmeasured variables $\boldsymbol{U}_{it}$, an assumption often called strict exogeneity \cite{callaway_treatment_2022}.
This can be viewed as a generalisation of the standard unconfoundeness assumption to the time-series cross-sectional setting, when unobserved confounding exist. 
\begin{assumption} \label{ass:EXOG}
(Strict exogeneity).
Conditional on observed covariates $\bm{X}_{it}$ and unmeasured variables $\boldsymbol{U}_{it}$, the intervention assignment mechanism does not depend on potential outcomes. For all $i$,
    \[
        \bar{\mathbf{A}}_{iT} \ci \left\{\{Y_{it}(\bar{\mathbf{a}}_t)\}_{\bar{\mathbf{a}}_t \in \bar{\mathcal{A}}_t}\right\}_{t=1}^T
        | \bar{\mathbf{X}}_{iT}, \bar{\mathbf{U}}_{iT}.
    \]
\end{assumption}

Figure \ref{fig:DAG} presents a causal directed acyclic graph (DAG) which is consistent with strict exogeneity (Assumption \ref{ass:EXOG}, \cite{xu_causal_2024}).
\begin{assumption} \label{ass:causalDAG}
(Causal DAG). The causal relationships between variables are as represented by the DAG in Figure \ref{fig:DAG}.
\end{assumption}

\tikzset{mynode/.style={minimum size=1.75cm, fill=gray!20, circle, align=center}}
\begin{figure}[ht]
\centering
\begin{tikzpicture}[node distance=1.5cm, font=\small]
    \node[mynode] (x0) at (0,0) {$\boldsymbol{X}_{t-1}$};
    \node[mynode] (x1) at (5,0) {$\boldsymbol{X}_{t}$};
    \node[mynode] (x2) at (10,0) {$\boldsymbol{X}_{t+1}$};
    \node[mynode] (y0) at (0,-3) {\small{$\boldsymbol{Y}_{t-1}$}};
    \node[mynode] (y1) at (5,-3) {\small{$\boldsymbol{Y}_{t}$}};
    \node[mynode] (y2) at (10,-3) {\small{$\boldsymbol{Y}_{t+1}$}};
    \node[mynode] (a0) at (0,-6) {$\boldsymbol{A}_{t-1}$};
    \node[mynode] (a1) at (5,-6) {$\boldsymbol{A}_{t}$};
    \node[mynode] (a2) at (10,-6) {$\boldsymbol{A}_{t+1}$};
    \node[mynode] (u0) at (0,-9) {$\boldsymbol{U}_{t-1}$};
    \node[mynode] (u1) at (5,-9) {$\boldsymbol{U}_{t}$};
    \node[mynode] (u2) at (10,-9) {$\boldsymbol{U}_{t+1}$};
    
    \draw[->] (x0) to (x1);
    \draw[->] (x1) to (x2);
    \draw[->] (a0) to (a1);
    \draw[->] (a1) to (a2);
    \draw[->] (u0) to (u1);
    \draw[->] (u1) to (u2);
    \draw[->] (x0) to (y0);
    \draw[->] (x0)[out=225] to (a0);
    \draw[->] (x1) to (y1);
    \draw[->] (x1)[out=225] to (a1);
    \draw[->] (x2) to (y2);
    \draw[->] (x2)[out=225] to (a2);
    \draw[->] (a0) to (y0);
    \draw[->] (a0) to (y1);
    \draw[->] (a0) to (y2);
    \draw[->] (a1) to (y1);
    \draw[->] (a1) to (y2);
    \draw[->] (a2) to (y2);
    \draw[->] (u0) to (a0);
    \draw[->] (u1) to (a1);
    \draw[->] (u2) to (a2);
    \draw[->] (u0)[out=115,in=240] to (y0);
    \draw[->] (u1)[out=115,in=240] to (y1);
    \draw[->] (u2)[out=115,in=240] to (y2);

\end{tikzpicture}
    \caption{Directed acyclic graph representing causal relationships between variables. The unit subscript $i$ is omitted to simplify the notation.}
    \label{fig:DAG}
\end{figure}
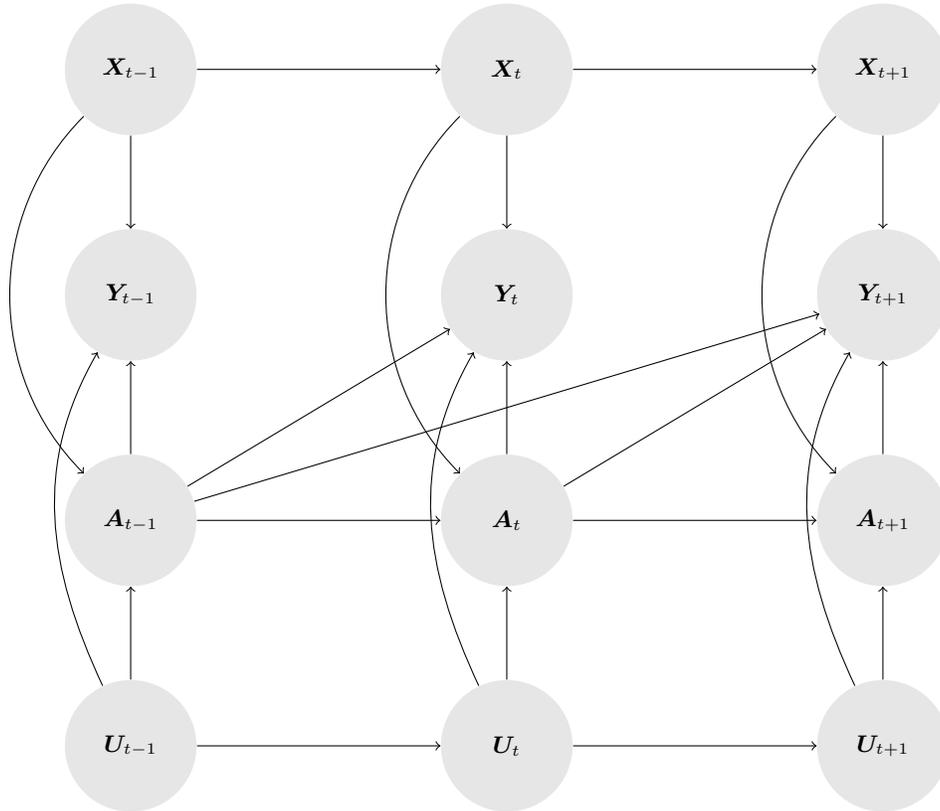

The key implications of the causal DAG in Figure \ref{fig:DAG} are that
(i) intervention assignment is not affected by past, current, or future outcomes (no arrow from any $Y_{ij}$ to $A_{it}$);
(ii) there is no feedback from past outcomes to current outcomes (no arrow from  $Y_{ij}$ to $Y_{it}$ for $j < t$);
and (iii) the observed and unobserved covariates are not affected by intervention assignment or outcomes (no arrows from any $Y_{ij}$ or any $A_{ij}$ to $X_{it}$ or $U_{it}$).

We now specify functional forms for the potential outcomes and the intervention assignment mechanism, which are consistent with strict exogeneity (Assumption \ref{ass:EXOG}) and the causal DAG (Assumption \ref{ass:causalDAG}).
Let $NegBin(a,b)$ denote the negative binomial distribution with mean $a$ and variance $a+a^2/b$.
\begin{assumption}\label{ass:functional1}
(Functional form for potential outcomes). For all $i$ and $t$,
\begin{equation}\label{eq:functional1}
    Y_{it}\left(\bar{\boldsymbol{a}}_{it}\right) \mid \boldsymbol{X}_{it}, \bm{U}_{it}, \bm{\eta}, \bm{\theta} \sim
    \begin{cases}
    NegBin(q^0_{it},\phi^0), & \text{if } \bar{\boldsymbol{a}}_{it}=\boldsymbol{0}_t\\
    NegBin(q^1_{it},\phi^1), & \text{if } \bar{\boldsymbol{a}}_{it}\neq\boldsymbol{0}_t
    \end{cases}
\end{equation}
for any $\bar{\boldsymbol{a}}_{t}\in\bar{\mathcal{A}}_t$, where 
\begin{eqnarray}
\label{eq:functional2} 
 \log\left(q_{it}^0\right)& = & f(\bm{U}_{it}) + \bm{\eta}^{\top} \boldsymbol{X}_{it}
 \\
\label{eq:functional3}
 \log\left(q_{it}^1\right)& = &   f(\bm{U}_{it}) + \bm{\eta}^{\top} \boldsymbol{X}_{it}+\psi(\bar{\boldsymbol{a}}_{it}, \boldsymbol{X}_{it};\bm{\theta}) 
 \\
\label{eq:fu}
  f(\bm{U}_{it})& = &   \kappa_i + \beta_t + \bm{\lambda}_i^{\top} \boldsymbol{V}_t
\end{eqnarray}
and 
\begin{equation}
\label{eq:functional4}
\psi(\bar{\boldsymbol{a}}_{it}, \boldsymbol{X}_{it},\bm{\theta})  = s\left(\sum_{j=1}^{t}a_{ij}\right)  +\theta_1\mathds{1}\{t_0<t<t_1\}+ \boldsymbol{\theta}_x \boldsymbol{X}_{it}
\end{equation}
for some smooth function $s(\cdot)$. 
\end{assumption}

In Equation \eqref{eq:functional1}, we use a different dispersion parameter $\phi^1$ for potential outcomes under intervention as we expect additional variability compared to potential untreated outcomes. 
Equations \eqref{eq:functional2}--\eqref{eq:functional3} are both variants of the factor analysis model. 
Factor models are popular in the field of causal inference as they allow to adjust for observed covariates (through $\boldsymbol{\eta}^\top\boldsymbol{X}_{it}$) and to control for unmeasured confounding (through $\kappa_i + \beta_t + \bm{\lambda}_i^{\top} \boldsymbol{V}_t$). 
Specifically, $\kappa_i$ can be interpreted as unit specific characteristics that stay constant across time, $\beta_t$ can be interpreted as shocks which affect all units equally, and $\boldsymbol{V}_t$ can be interpreted as common shocks which affect all units differently through $\lambda_i$.
Although both $\kappa_i$ and $\beta_t$ can be absorbed by the term $\boldsymbol{\lambda}_i^\top\boldsymbol{V}_t$, we find it useful to separate out those additional terms to highlight that the model we use is a generalisation of the linear difference-in-differences (DD) model often used in the field of causal inference for cross-sectional time-series data.

The term $\psi(\bar{\boldsymbol{a}}_{t}, \boldsymbol{X}_{it};\bm{\theta})$ in Equation \eqref{eq:functional3} models how the intervention effect depends on the exact nature of the intervention (i.e.\ numbers of peers at each time in the past and present), the presence of a lockdown and the observed covariates $\boldsymbol{X}_{it}$. 
We model the possibly non-linear relationship between cumulative exposure up to time $t$, $\sum_{j=1}^{t}a_{j}$, and the outcome using B-splines \cite{perperoglou_review_2019} i.e.\ we assume
\begin{equation}\nonumber
s\left(\sum_{j=1}^{t}a_{j}\right) =
\sum_{b=1}^{p+b^*} w_b B_{b} \left(\sum_{j=1}^{t} a_{j}\right) ,
\end{equation}
where $p$ is the degree of the B-spline, $b^*$ is the number of knots, $B_{b} \left(\sum_{j=1}^{t} a_{j}\right)$ represents the basis functions and $w_b$ are the basis coefficients. 
Here we set $p=3$ (cubic spline) and $b^\ast=3$. 
Splines have previously been considered for modelling the effect of time in intervention in stepped-wedge trials \cite{kenny_analysis_2022, wu_advancing_2024}. 

We now specify the functional form for the assignment of the intervention path.

\begin{assumption}
    \label{ass:functional2}
(Functional form for intervention assignment). For all $i$ and $t$,
\begin{equation}\label{eq:functional5}
A_{it} = 
\begin{cases}
0,& t<t_\mathrm{min}
\\
A_{i,t-1} + M_{it}, & t\geq t_\mathrm{min}
\end{cases},
\end{equation}
where $t_\mathrm{min}$ is the earliest time point at which a peer could be introduced, $M_{it}$ is the total number of peers recruited by unit $i$ at time $t$, and 
\begin{equation} \nonumber
    M_{it} \mid  \boldsymbol{X}_{it}, \boldsymbol{U}_{it}, \delta_0, \boldsymbol\delta_x \sim Pois(\mu_{it})
\end{equation}
\begin{equation}\nonumber
\log(\mu_{it}) = \delta_0 + g(\boldsymbol{U}_{it}) +\boldsymbol\delta_x^\top\boldsymbol{X}_{it}.
\end{equation}
\begin{equation} \nonumber
    g(\boldsymbol{U}_{it}) = \delta_\kappa\kappa_i + \boldsymbol{\delta}_\lambda ^\top\boldsymbol{\lambda}_i
\end{equation}
\end{assumption}
Assumption \ref{ass:functional2} encodes our prior belief that in an observational study, assignment of the intervention may depend on a unit's characteristics. 
This is a key difference with randomised studies (for which $\mu_{it}$ would be constant across units). 
Note that Assumption \ref{ass:functional2} is consistent with irreversibility of intervention (Assumption \ref{ass:NonIrreversibility}).
In addition, note that $f(\boldsymbol{U}_{it})$ in Assumption \ref{ass:functional1} and $g(\boldsymbol{U}_{it})$ in Assumption \ref{ass:functional2} share parameters $\kappa_i$ and $\bm\lambda_i$, which govern the confounding caused by $\bm{U}_{it}$.

We now make some remarks regarding our model. 
First, when there are no covariates $\boldsymbol{X}_{it}$, as in our application, one can write the estimands in Equation \eqref{eq:psi} using \eqref{eq:functional2}--\eqref{eq:functional3} and integrating over $\{\kappa_i, \boldsymbol{\lambda}_i\}$ as:
\begin{multline}\nonumber
\omega_t(\bar{\boldsymbol{a}}_t) =
\frac{\int\int E\left[Y_{it}(\bar{\boldsymbol{a}}_t)\mid k_i, \boldsymbol\lambda_i\right]\pi\left(\kappa_i,\boldsymbol \lambda_i \right)d\kappa_id\boldsymbol\lambda_i}
{\int\int E\left[Y_{it}(\boldsymbol{0}_t)\mid k_i,\boldsymbol\lambda_i \right]\pi\left(\kappa_i,\boldsymbol\lambda_i\right)d\kappa_id\boldsymbol\lambda_i}
\\=\exp\left(\psi(\bar{\boldsymbol{a}}_{it})\right)=\exp\left(s\left(\sum_{j=1}^{t}a_{j}\right)\right)\exp\left(\theta_1\mathds{1}\{t_0<t<t_1\}\right).
\end{multline}
Thus, the $\omega_t(\bar{\boldsymbol{a}}_t)$ depend only cumulative intervention intensity and the time $t$. 
It is worth noting that when there are covariates, obtaining $\omega_t(\bar{\boldsymbol{a}}_t)$ is hard because we need to integrate over the $\boldsymbol{X}_{it}$. 
In such cases, it is common to use the CATEs $\omega_t(\bar{\boldsymbol{a}}_t,\boldsymbol{x}_{it})$ to obtain the so-called mixed average intervention effects
\begin{equation}\nonumber
    \tau_m = \frac{1}{\tilde{N}_1}\sum_{i,t:A_{it} > 0} \omega_t(\bar{\boldsymbol{a}}_t,\boldsymbol{X}_{it}),
\end{equation}
where $\tilde{N}_1=\sum_{i,t} \mathds{1}\{i,t:A_{it} > 0\}$, i.e.\ to integrate over the empirical distribution of the covariates \cite{ding_causal_2018}. 
The CATEs are
\begin{multline} \nonumber
\omega_t(\bar{\boldsymbol{a}}_t, \boldsymbol{x}_{it}) =
\frac{\int\int E\left[Y_{it}(\bar{\boldsymbol{a}}_t)\mid \boldsymbol{x}_{it},k_i,\boldsymbol\lambda_i\right]\pi\left(\kappa_i,\boldsymbol\lambda_i\right)d\kappa_id\boldsymbol\lambda_i}
{\int\int E\left[Y_{it}(\boldsymbol{0}_t)\mid \boldsymbol{x}_{it},k_i,\boldsymbol\lambda_i\right]\pi\left(\kappa_i,\boldsymbol\lambda_i\right)d\kappa_id\boldsymbol\lambda_i}
\\=\exp\left(s\left(\sum_{j=1}^{t}a_{ij}\right)\right)\exp\left(\theta_1\mathds{1}\{t_0<t<t_1\}\right)\exp\left(\boldsymbol\theta_x\boldsymbol{x}_{it}\right).
\end{multline}

Second, it is possible to provide causal interpretations for some of the parameters in \eqref{eq:functional4}. 
For example, in Supplementary Material A, we provide an interpretation for $\exp\left(\theta_1\right)$ as a rate ratio between two potential outcomes. 
However, the contrasts these parameters address are not often considered in the causal inference literature. 
Third, it is straightforward to modify our approach to accommodate outcomes of different type (e.g.\ Gaussian or binomial) and interventions with continuous intensities (e.g.\ dose), by modifying the models in Assumptions \ref{ass:functional1} and \ref{ass:functional2}, respectively.

Fourth, the functional form for intervention-free potential outcomes $Y_{it}(\bm0_t)$ nested in Assumption \ref{ass:functional1} together with Assumptions \ref{ass:NoInterference} through \ref{ass:causalDAG} can be used estimate a counterfactual imputation factor analysis model \cite{nethery_integrated_2023, samartsidis_bayesian_2024}.
To estimate this model, we only employ pre-intervention outcomes, thus, we will refer to this model as the \textit{pre-intervention outcome model} from now on.
As the parameters $\kappa_i$ and $\bm\lambda_i$ are shared between $f(\boldsymbol{U}_{it})$ in Assumption \ref{ass:functional1} and $g(\boldsymbol{U}_{it})$ in Assumption \ref{ass:functional2}, the pre-intervention outcome model is clearly disregarding some parts of the data that are informative about parameters in the model for the intervention-free potential outcomes.

\subsection{Bayesian estimation}
\label{subsec:bayes}
We now describe how to perform inference on the estimands defined in Section \ref{subsec:causal} under the Bayesian paradigm. 
Define $\mathrm{data} = \left\{ \left\{ A_{it}, Y_{it}, \boldsymbol{X}_{it} \right\}_{t=1}^T \right\}_{i=1}^N$.
Further, assume exchangeability across units (see e.g. \cite{ben-michael_estimating_2023, pang_bayesian_2022}). 
Let $\boldsymbol\Xi$ be a generic parameter governing the joint distribution of variables shown in Figure \ref{fig:DAG}.
Using exchangeability and Assumptions \ref{ass:consistency} and  \ref{ass:causalDAG}, we obtain the likelihood as 
\begin{equation}\label{eq:lik2}
\mathbb{P}\left(\mathrm{data}\mid\boldsymbol\Xi\right) =
\prod_{i=1}^{N} \left[ \prod_{t=1}^{T} \mathbb{P}\left(Y_{it} \mid \bar{\boldsymbol{A}}_{it}, \boldsymbol{X}_{it}, \boldsymbol\Xi\right)
\mathbb{P}\left(A_{it}\mid\bar{\boldsymbol{A}}_{i,t-1}, \boldsymbol{X}_{it}, \boldsymbol\Xi\right)\right] 
\times \mathbb{P}\left(\bar{\boldsymbol{X}}_{iT} \mid \boldsymbol\Xi\right),
\end{equation}
where the conditional distributions inside the square brackets are defined in Assumptions \ref{ass:functional1} and \ref{ass:functional2}, and $\{\{\kappa_i , \bm{\lambda}_i \}_{i=1}^N, \{\beta_t, \boldsymbol{V}_t \}_{t=1}^T, \delta_\kappa, \bm\delta_\lambda\} \subset \bm\Xi$ are treated as model parameters since they are unobserved quantities. 
Finally, we assume that the right-hand side of \eqref{eq:lik2} can be rewritten as 
\begin{equation}\label{eq:lik3}
\prod_{i=1}^{N}\left[\prod_{t=1}^{T}\mathbb{P}\left(Y_{it} \mid \bar{\boldsymbol{A}}_{it}, \boldsymbol{X}_{it}, \boldsymbol\Theta \right)
\mathbb{P}\left(A_{it}\mid\bar{\boldsymbol{A}}_{i,t-1},\boldsymbol{X}_{it},\boldsymbol\Theta\right)\right] \times
\mathbb{P}\left(\bar{\boldsymbol{X}}_{iT} \mid \boldsymbol\Xi_x\right),
\end{equation}
where $(\boldsymbol\Theta, \bm\Xi_x)= \bm\Xi$. 
$\bm\Theta$ is governing the distribution of the outcomes and the assignment mechanism; $\boldsymbol\Xi_x$ is governing the distribution of observed covariates; and $\mathbb{P}\left(\boldsymbol\Theta,\boldsymbol\Xi_x\right)=\mathbb{P}\left(\boldsymbol\Theta\right)\mathbb{P}\left(\boldsymbol\Xi_x\right)$ (independent priors). 
This allows us to exclude likelihood contributions $\mathbb{P}\left(\bar{\boldsymbol{X}}_{iT}\mid\bm\Xi_x\right)$ from further analyses.
This is advantageous since posterior computations are sped up and there is no need to specify a functional form for the distribution of covariates. 

The set of parameters of interest $\boldsymbol\Theta$ is
\begin{equation}\nonumber
\boldsymbol\Theta=\left\{ \{\kappa_i, \bm{\lambda}_i \}_{i=1}^n, \{\beta_t, \boldsymbol{V}_t \}_{t=1}^T, \boldsymbol\eta, \phi^0, \phi^1, \theta_1, \boldsymbol\theta_x, \boldsymbol{w},\delta_0, \delta_\kappa, \boldsymbol\delta_\lambda, \boldsymbol\delta_x \right\}
\end{equation}
with posterior
\begin{multline}\label{eq:posterior}
\mathbb{P}\left(\boldsymbol\Theta\mid\mathrm{data}\right)\propto
\mathbb{P}\left(\boldsymbol\Theta\right) \\
\prod_{i=1}^{N}
\left[\prod_{t<g_i}NegBin(Y_{it};q_{it}^0,\phi^0)
\prod_{t\geq g_i} NegBin(Y_{it};q_{it}^1,\phi^1) 
\prod_{t \geq t_{min}}Pois(A_{it}-A_{i,t-1}; \mu_{it}) \right],
\end{multline}
following when combining \eqref{eq:lik3} with Assumptions \ref{ass:functional1} and \ref{ass:functional2}.

The inclusion of the intervention assignment model in \eqref{eq:lik2}, \eqref{eq:lik3}, and \eqref{eq:posterior} is one of the key differences of our approach compared to existing Bayesian counterfactual imputation methods. 
The majority of methods approaches ignore the contribution to the likelihood of the terms associated to intervention assignment (e.g.\ $\mathbb{P}\left(A_{it}\mid\bar{\boldsymbol{A}}_{i,t-1},\boldsymbol{X}_{it},\Theta\right)$ in \eqref{eq:lik2}). 
This approach is valid when $\{\boldsymbol\lambda_i, \kappa_i\}_{i=1}^n$ do not affect $\{\bar{\boldsymbol{A}}_{iT}\}_{i=1}^n$ (i.e.\ there is no confounding through unmeasured variables $\bm{U}_{it}$). 
When this is not true, the posterior obtained by ignoring the likelihood terms associated to the intervention assignment is a cut posterior \cite{plummer_cuts_2015}. 
Although cut posteriors have good properties when both $n$ and $T$ are large \cite{Pompe2021}, their properties when applied to small datasets are less clear. 
In Supplementary Material C, we demonstrate through a simulation study that using a cut posterior approach may lead to poor coverage of credible intervals. 
Discarding likelihood terms associated to the intervention assignment may also lead to a loss of efficiency when $A_{it}$ are very informative regarding $\boldsymbol\lambda_i$ and $\kappa_i$.

We conclude our Bayesian model specification by specifying prior distributions for the elements of $\boldsymbol{\Theta}$. 
Most of them are assigned vague prior distributions. 
For $i=1, \ldots, n$, we assume that $\kappa_i\sim Normal(\mu = 0, \sigma = 50)$, where $\mu$ is the mean and $\sigma$ is the standard deviation of the normal distribution. 
Most of the remaining scalar parameters ($\{\beta_t\}$, $\theta_1$, $\theta_2$, $\delta_0$ and $\delta_\kappa$) are given $Normal(0,10)$ priors. 
Let $\mathscr{I}_D$ be the $D\times D$ identity matrix. 
We assign $\mathcal{N}(\boldsymbol{0}_D, 10\mathscr{I}_D)$ priors to the vector-valued parameters of the model ($\{V_t\}$, $\boldsymbol{w}$, $\boldsymbol{\theta}_x$, $\boldsymbol{\delta}_\lambda$, and $\boldsymbol{\delta}_x$) where $D$ denotes the dimension of each respective vector.
For $\{\lambda_i\}$ we assign $\mathcal{N}(\boldsymbol{0}_D, 50\mathscr{I}_D)$.
For dispersion parameters, we follow Gelman et al. \cite{gelman_prior_2025} and specify a prior on $\frac{1}{\sqrt{\phi^0}}$ and $\frac{1}{\sqrt{\phi^1}}$. 
Specifically, we let $\frac{1}{\sqrt{\phi^0}} \sim Normal(0, \sigma_0)$ and $\frac{1}{\sqrt{\phi^1}} \sim Normal(0, \sigma_1 )$ and let $\bar{Y}_{it,0} = \frac{1}{n_0} \sum_{it:A_{it}=0} Y_{it}$ and $\bar{Y}_{it,1} = \frac{1}{n_1} \sum_{it:A_{it}>0} Y_{it}$, where $n_0$ and $n_1$ are the number of exposed and unexposed observations, respectively.
$\sigma_0$ and $\sigma_1$ can be chosen such that the prior probabilities of $1+\frac{\bar{Y}_{it,0}}{\phi_0}$ and $1+\frac{\bar{Y}_{it,1}}{\phi_1}$ being greater than three are about 0.05.

$\psi(\bar{\boldsymbol{a}}_t)$ and $\psi(\bar{\boldsymbol{a}}_t,\boldsymbol{x}_{it})$ are functions of some model parameters, namely $\{\theta_1,\{w_b\},\boldsymbol\theta_x\}$.
Thus, inference is straightforward through the posterior distribution of parameters.

The ITEs $\tau_{it}$ (and associated estimands) cannot be obtained directly from the posterior \eqref{eq:posterior} as they rely on the missing potential outcomes $Y_{it}(\boldsymbol{0}_t)$ for $i,t: A_{it}>0$. 
Let $\mathcal{Y}=\{Y_{it}(\boldsymbol{0}_t)\}_{i,t:A_{it}>0}$ be the set of missing potential outcomes. 
It can be shown (see Supplementary Material A) that 
\begin{equation}\label{eq:predictive}
\mathbb{P}\left(\mathcal{Y}\mid\mathrm{data}\right) = \int\prod_{i,t:A_{it}>0}\left[\mathbb{P}\left(Y_{it}(\boldsymbol{0}_t)\mid Y_{it}(\bar{\boldsymbol{A}}_t), \boldsymbol{X}_{it},\boldsymbol\Theta\right)\right]\mathbb{P}\left(\boldsymbol\Theta\mid\mathrm{data}\right)d\boldsymbol\Theta.
\end{equation}
We draw $\mathcal{Y}$ from this posterior predictive distribution. 
However, it is not possible to draw from $\mathbb{P}\left(Y_{it}(\boldsymbol{0}_t)\mid Y_{it}(\bar{\boldsymbol{A}}_t), \boldsymbol{X}_{it},\boldsymbol\Theta\right)$ unless some assumptions are made regarding the joint distribution of $Y_{it}(\boldsymbol{0}_t)$ and $Y_{it}(\bar{\boldsymbol{a}}_t)$ ($\bar{\boldsymbol{a}}_t\neq\boldsymbol{0}_t$). 
Here, we assume that for any $\bar{\boldsymbol{a}}_t\neq\boldsymbol{0}_t$, the joint cumulative distribution function (cdf) $H$ of $Y_{it}(\boldsymbol{0}_t)$ and $Y_{it}(\bar{\boldsymbol{a}}_t)$ conditional on the covariates $\boldsymbol{X}_{it}$ and $\boldsymbol{\Theta}$ is described by a bivariate Gaussian copula (see e.g.\ \cite{hoff_extending_2007}) with correlation parameter $\rho$, i.e.\
\begin{multline}\label{eq:copula}
    H\left(Y_{it}(\boldsymbol{0}_t),Y_{it}(\bar{\boldsymbol{a}}_t) \mid \boldsymbol{X}_{it},\boldsymbol\Theta,\rho\right)= \\
    \Phi_2\left(\Phi^{-1}\left(\Psi\left(Y_{it}(\boldsymbol{0}_t); q_{it}^{0},\phi^{0}\right)\right),\Phi^{-1}\left(\Psi\left(Y_{it}(\bar{\boldsymbol{a}}_t);q_{it}^{1},\phi^{1}\right)\right);\boldsymbol{0}_2,\mathscr{R}\right),
\end{multline}
where $\Phi_2(\cdot,\cdot;m,R)$ is the cdf of a bivariate Gaussian distribution with mean vector $m$ and covariance matrix $R$, $\Phi(\cdot)$ is the cdf of the standard Gaussian distribution, $\Psi(\cdot;q,\phi)$ is the cumulative density function of the $NegBin(q,\phi)$ (see Assumption \ref{ass:functional1}), and $\mathscr{R}=\left[\begin{array}{cc}1&\rho\\ \rho& 1\end{array}\right]$.

Using Equation \eqref{eq:copula} allows us to draw the missing potential outcomes from the posterior predictive as \cite{alexopoulos_bayesian_2021}
\[
Y_{it}\left(\boldsymbol{0}_t\right)=\mathrm{min}\left\{y:\Psi(y;q_{it}^0,\phi^0)\geq \Phi(z_{it}^0) \right\}, 
\]
where 
\begin{eqnarray}
\nonumber
z_{it}^0 &\sim  & Normal( \rho z_{it}^1,1-\rho^2),
\\ \nonumber
z_{it}^1 & = & \Phi^{-1}(u_{it}),
\\ \nonumber
u_{it} & \sim & Uniform\left( 
\Psi(Y_{it}-1;q^1_{it},\phi^1),
\Psi(Y_{it};q^1_{it},\phi^1) \right).
\end{eqnarray}

As only one of $Y_{it}(\boldsymbol{0}_t)$ and $Y_{it}(\bar{\boldsymbol{a}}_t)$ can be observed, it is not possible to estimate $\rho$. 
Thus, it is recommended that sensitivity of findings to the choice of this parameter is always checked \cite{ding_causal_2018}. 
A sensitivity analysis is computationally cheap to conduct as it does not require evaluating a different posterior each time. 
An alternative to sensitivity analysis is to integrate over the uncertainty on $\rho$, by assigning a vague prior to this parameter, e.g.\ a $Uniform(-1,1)$. 
We use both approaches on the real data. 
This is another important difference of our approach compared to existing methods for Bayesian causal inference using time-series data.
To the best of our knowledge all available methods side-step the issue by making convenient assumptions about the joint distribution of the potential outcomes, most commonly independence of potential outcomes conditional on the covariates and model parameters ($\rho=0$).

The posterior distributions of all estimands defined in Section \ref{subsec:causal} can be obtained directly from either the posterior \eqref{eq:posterior} or the posterior predictive \eqref{eq:predictive}, using the appropriate transformations. 
Point estimates are obtained as the posterior means, and 95\% credible intervals (CrIs) are constructed using the 2.5\% and 97.5\% quantiles of the posterior distribution.

\subsection{Practical implementation}
\label{subsec:computations}

The main difficulty in performing inference regarding the causal effects as described in Section \ref{subsec:bayes} is to evaluate the posterior \eqref{eq:posterior}, which is analytically intractable. 
We resort to Markov chain Monte Carlo (MCMC) to draw samples from it. 
Specifically, samples are collected using the No-U-Turn sampler (NUTS) \cite{hoffman_no-u-turn_2014}. 
NUTS is an adaptive variant of the Hamiltonian Monte Carlo algorithm for sampling from high-dimensional posteriors, which requires no tuning from the user.
Our method is implemented in the statistical language \texttt{R} using the \texttt{rstan} package \cite{stan_development_team_rstan_2020, r_core_team_r_2024}.
Sampling the latent factors $\boldsymbol{V}_t$ and loadings $\bm{\lambda}_i$ is challenging as those parameters are not identified without further constraints (see e.g.\ \cite{zhao_bayesian_2016}), which can slow down MCMC mixing. 
In our experiments, we noticed that although the mixing of these parameters was poor, this did not affect the mixing of identifiable parameters such as $q_{it}^{1},\phi^1,q_{it}^0$ and $\phi^0$. 
Since we are not interested in performing inference on $\boldsymbol{V}_t$ or $\bm{\lambda}_i$ individually (although we are interested in their product), they are left unidentified.  

Another challenge is that the dimension $h$ of latent variables $\boldsymbol\lambda_i$ cannot be known in advance. 
Cross-validation (CV) can be used to select the number of factors \cite{xu_generalized_2017}.
The factor model developed in this paper employs the leave-one-out-CV algorithm described in Algorithm 1 in Supplementary Material B.
The basic idea of CV is to hold back one randomly selected observation from each exposed unit.
Then, the Bayesian factor model is fitted for various choices of the number $h$, and the posterior predictive distribution of the missing observations is obtained for each $h$.
The algorithm chooses the model that, on average, predicts the missing data most accurately.

To assess prediction accuracy, the mean squared prediction error (MSPE) is used.
The MSPE is the mean squared difference between the observed outcomes and draws from the posterior predictive distribution for that outcome.
The MSPE strongly penalises large prediction errors.
As an additional measure of prediction accuracy, we use the interval score (IS).
The IS is the width of the 95\%-CrI plus a term that penalises if the observed value lies outside the 95\%-CrI \cite{gneiting_strictly_2007}.
The advantage of the IS over the MSPE is that it takes into account the width of the 95\%-CrI.
If MSPE and IS lead to different results, the researcher should further investigate.
For example, there might be one outlier observation driving the results.

\section{Evaluation of the peers intervention}
\label{sec:Results}

We now evaluate the effect of the peers intervention on case-finding of DAA-eligible individuals using the methods introduced in the preceding section.
In Section \ref{subsec:implementation_details}, we lay out the implementation details. 
Results are presented and discussed in Section \ref{subsec:results}.

\subsection{Implementation details} \label{subsec:implementation_details}

We choose the number of factors using the CV approach described in Algorithm 1 in Supplementary Material B. 
Table \ref{table:CVresults} shows results obtained for $h=0,\ldots,5$. 
When $h=0$, the model that we fit is a variant of the DD model. 
We see that the MSE and IS were both minimised for the model with one factor. 
We thus set $h=1$ for subsequent analyses.

\begin{table}
\caption{Cross-validation results for number of factors using Algorithm 1 (Supplementary Material B) with $M=50$ data sets.
Abbreviations: DD: Difference-in-Differences; MSPE: Mean Squared Prediction Error.}
\label{table:CVresults}
\begin{tabular}{@{}lllllll@{}}
\hline
& \multicolumn{6}{c}{Number of Factors} \\
\cline{2-7}
Metric & 0 (DD) & 1 & 2 & 3 & 4 & 5 \\
\hline
MSPE           & 399.87 & \textbf{364.43} & 380.02 & 395.22 & 405.78 & 405.81 \\
Interval score & 66.61  & \textbf{65.08}  & 66.97  & 67.31  & 67.53  & 65.51  \\
\hline
\end{tabular}
\end{table}

We run two MCMC chains each for 100,000 iterations, discarding the first 50,000 as burn-in and thinning the remaining draws at every 5 iterations to obtain a total of 20,000 draws from the posterior. 
Prior distributions are set as in Section \ref{subsec:bayes}. 
The values of $\sigma_0$ and $\sigma_1$ are 0.11 and 0.1165, respectively.
We consider nine different priors for $\rho$, including point mass priors at 1, 0.75, 0.5, 0 and -1, and uniform priors over the intervals $[0.75,1]$, $[0.5,1]$, $[0,1]$ and $[-1,1]$. 
Convergence is assessed using the $\hat{R}$ statistic and by visual inspection of posterior trace plots for some identifiable parameters \cite{vehtari_rank-normalization_2021}. 
Some trace plots are shown in Supplementary Material D. 
We see that both chains have converged to the same stationary distribution, and that mixing is satisfactory.

To check the sensitivity of the results to the choice of prior distributions for all parameters except $\rho$, we repeat the analysis a set of vague priors. 
Namely, we increase the standard deviation of Gaussian priors by a factor of 10, except for $\phi_0$ and $\phi_1$ (to avoid attributing all variability in the data to noise).
The results from this sensitivity analysis are almost identical to those obtained using the original priors and are therefore not discussed further.

We perform two additional analyses of the data using alternative approaches, to compare with our fully Bayesian method. 
First, we implement the \textit{pre-intervention outcome model}.
Without using post-intervention data it is not possible to implement our proposed sensitivity analysis for assumptions regarding the joint distribution of potential outcomes.
Thus, we are forced to assume that the potential outcomes are independent ($\rho=0$) when using the pre-intervention outcome model.
Second, we implement a cut posterior approach that discards likelihood terms associated to the intervention assignment only (henceforth \textit{outcome model}). 
We use the same MCMC specifications and prior distributions as in our fully Bayesian model.

\subsection{Results} \label{subsec:results}

\begin{figure}[t]
    \centering
        \includegraphics[scale=0.65]{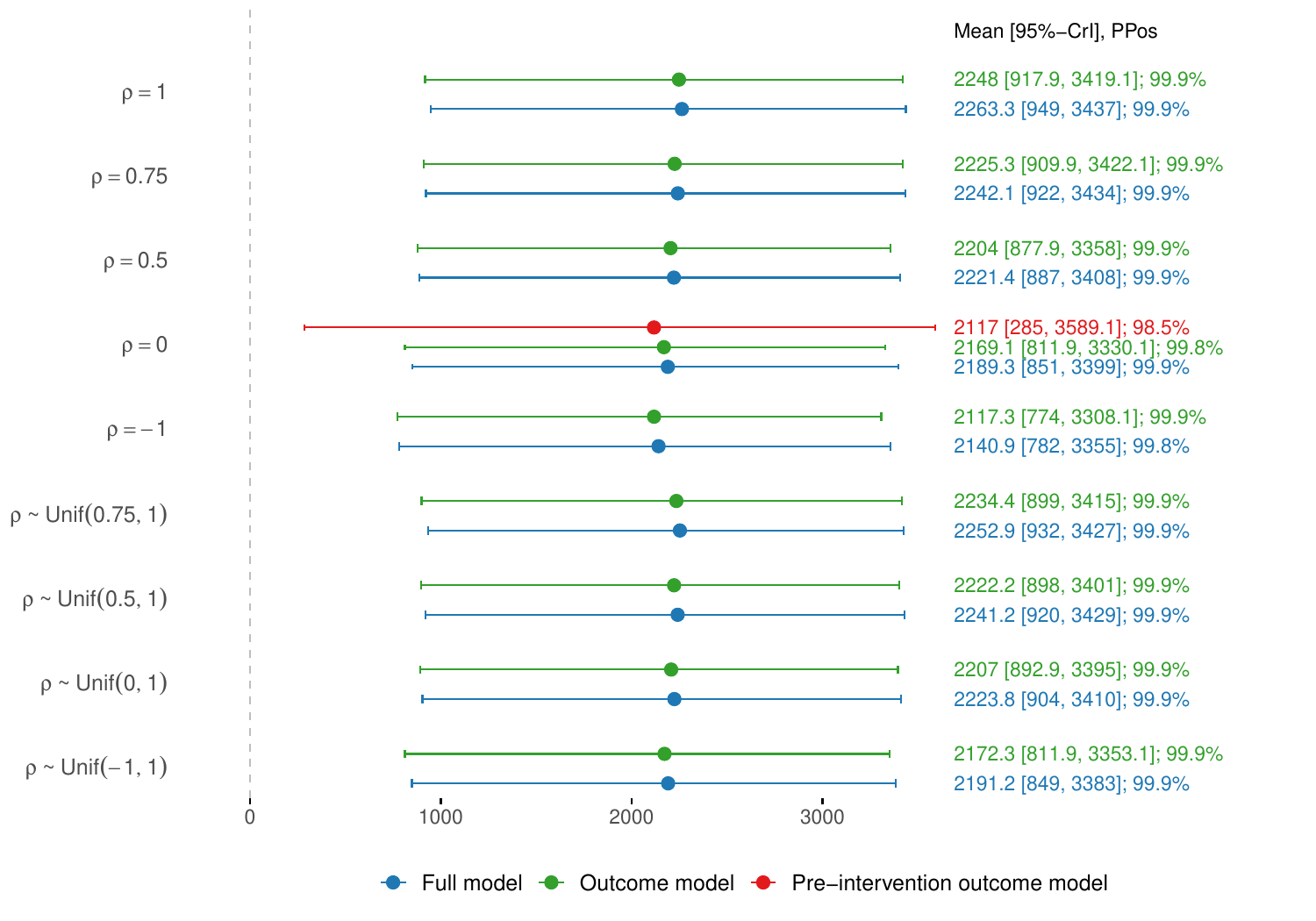}
    \caption[Estimated cumulative intervention effects]
    {
    Estimated cumulative number of treatment eligible Hepatitis-C-virus-infected individuals identified due to the peers intervention.
    Abbreviations: 95\%-CrI: 95\% credible interval; PPos: Posterior probability of a positive intervention effect.
    The cumulative effects were taken across the whole study period and all Operation Delivery Networks.
    $\rho$ is the assumed correlation between the potential outcomes using the Gaussian copula approach.
    The \textit{full model} uses all available data (pre- and post-intervention outcomes and intervention assignment), the \textit{outcome model} discards intervention assignment, and the \textit{pre-intervention outcome model} discards post-intervention outcomes and assignment mechanism.
    }
    \label{fig:Results_rho}
\end{figure}

First, we present results that summarise the evidence for an overall intervention effect. 
Figure \ref{fig:Results_rho} shows the point estimates and 95\%-CrIs for the cumulative intervention effect $\tau$, under the different prior specifications for $\rho$ (blue dots and bands). 
We see that the point estimates are not sensitive to the choice of prior for $\rho$.
All nine point estimates suggest a large positive intervention effect. 
For example, for $\rho=1$ and $\rho=0$, we estimate that 2248.0 and 2189.3 additional referrals, respectively, were made thanks to the contribution of the peers. 
Further, all nine priors yield a very high (at least 99\%) probability of a positive intervention effect (PPos). 
In Supplementary Material E, we present posterior summaries for the percentage increase in case finding thanks to the peers ($\chi$), under the different priors for $\rho$. 
For $\rho=1$ and $\rho=0$, we estimate that the peers intervention increased total case-finding by 17.3\% (95\% CrI [6.5, 28.3]) and 16.6\% (95\% CrI [5.8, 27.9]), respectively.

\begin{figure}[t]
    \centering

    \includegraphics[width=0.4\linewidth]{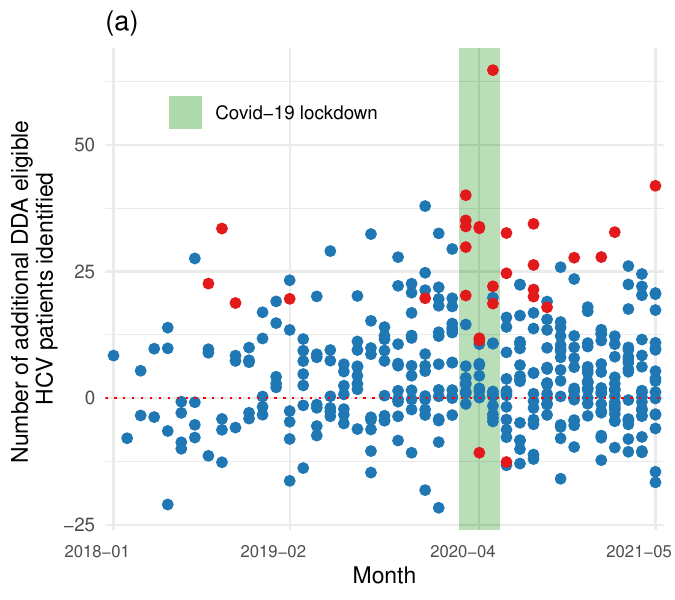}
    \includegraphics[width=0.4\linewidth]{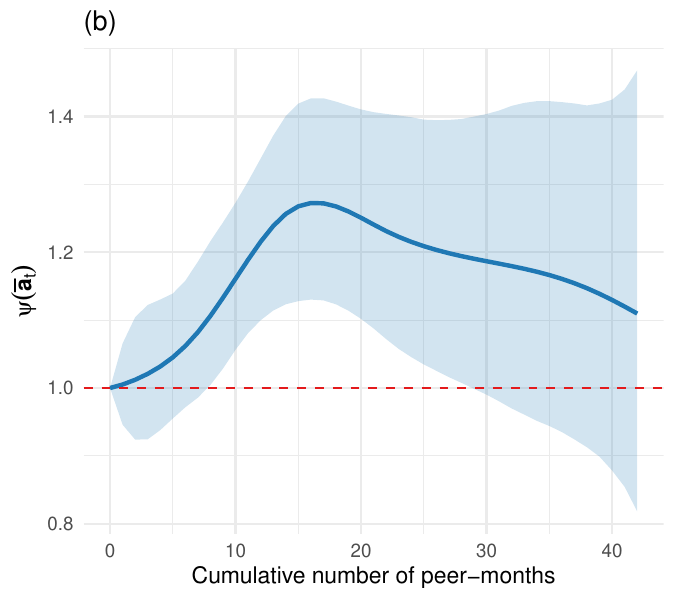}

    \includegraphics[width=0.4\linewidth]{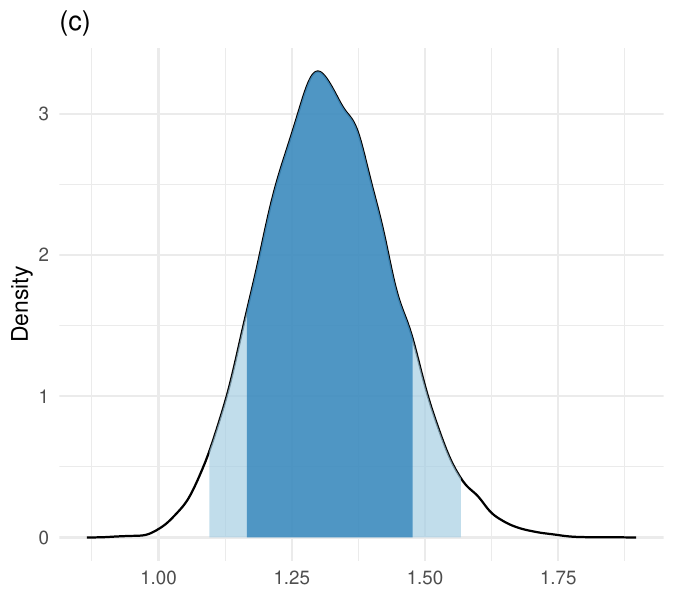}
    \includegraphics[width=0.4\linewidth]{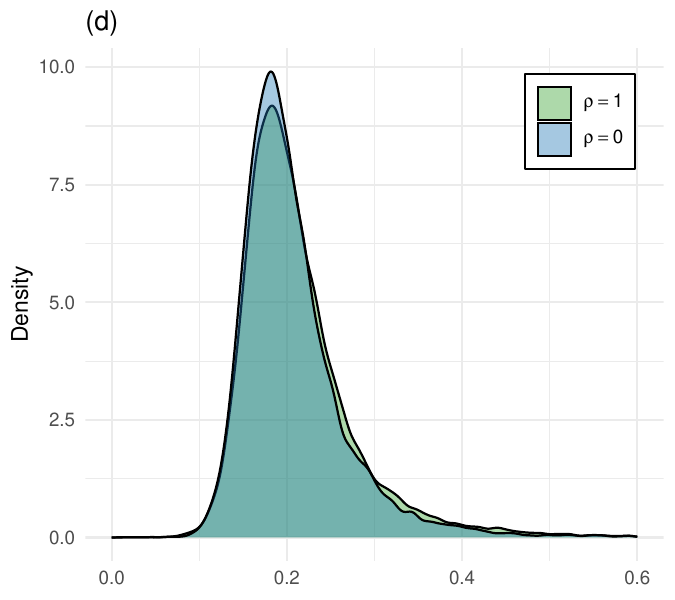}

    \caption{
    Additional results graphs.
    (a) Point estimates (posterior mean) of $\tau_{it}$ for $\rho=0$ (394 in total). Red dots signal that the 95\% credible interval did not include 0, while blue dots signal that it did. 
    (b) Point estimates and 95\% CrIs for $\exp\left(s\left(\sum_{j=1}^{t}a_{j}\right)\right)$. 
    (c) Posterior distribution of $\exp(\theta_1)$. The dark blue indicates the 80\% CrI and the light blue indicates the 95\% CrI.
    (d) Posterior distribution of the share of additional Hepatitis-C-virus-infected individuals identified during the COVID-19 lockdown relative to the overall number of additional Hepatitis-C-virus-infected individuals identified ($\tau_c/\tau$). 
    }
    \label{fig:add_results}
\end{figure}

\begin{figure}[ht!]
\centering
\resizebox{.85\linewidth}{!}{%
\begin{tabular}{c c c}
    & \textbf{$\rho=1$} & \textbf{$\rho=0$} \\

    \rotatebox{90}{\textbf{Kent}} &
    \includegraphics[width=0.4\linewidth]{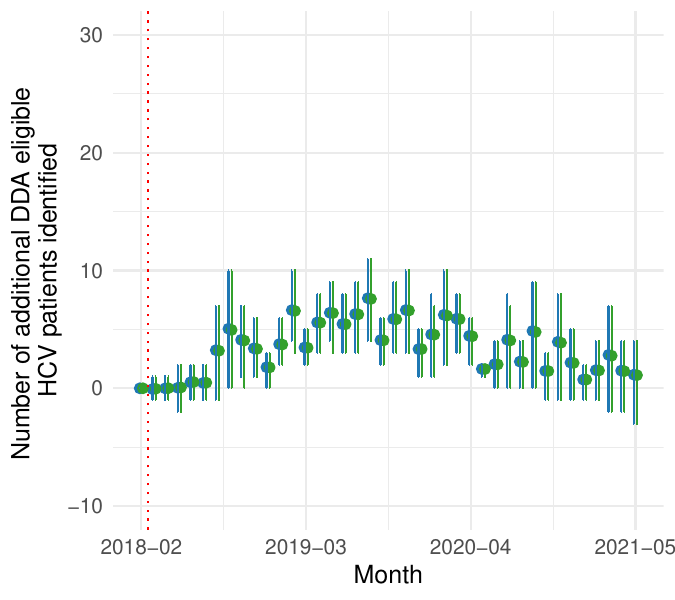} &
    \includegraphics[width=0.4\linewidth]{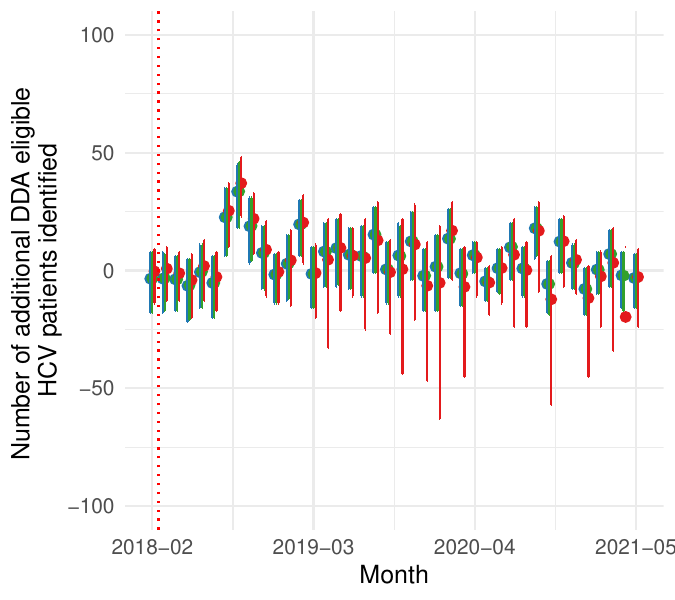} \\

    \rotatebox{90}{\textbf{Merseys. \& Che.}} &
    \includegraphics[width=0.4\linewidth]{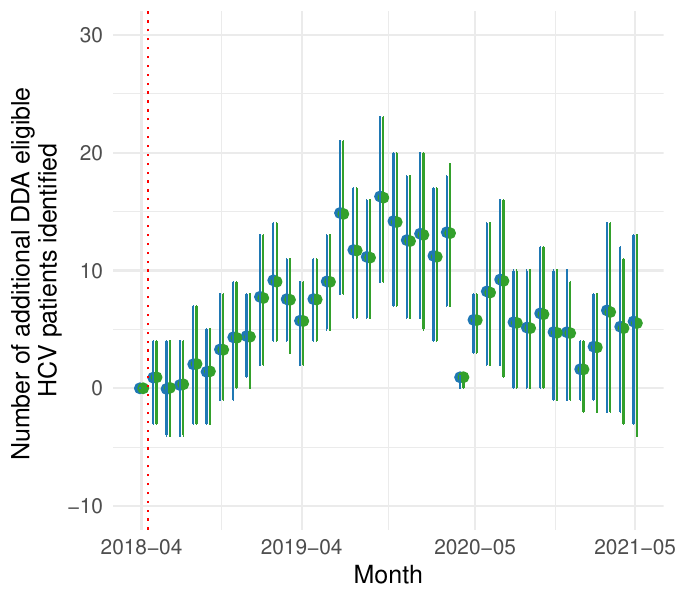} &
    \includegraphics[width=0.4\linewidth]{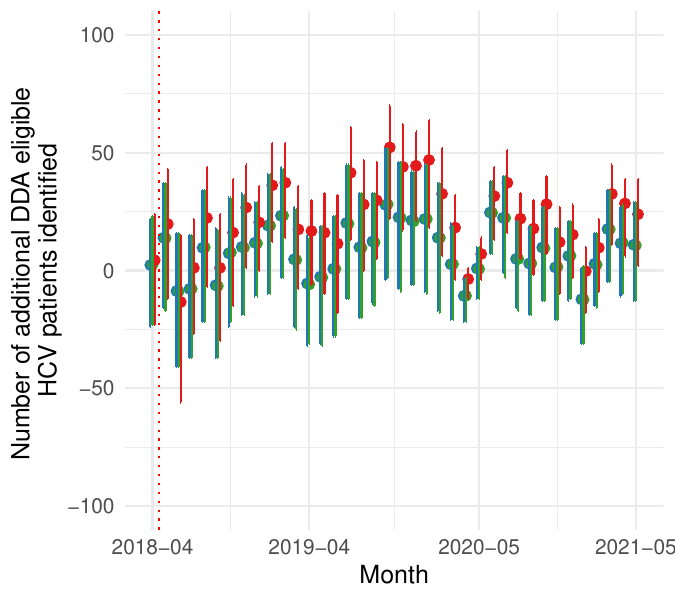} \\

    \rotatebox{90}{\textbf{South Yorkshire}} &
    \includegraphics[width=0.4\linewidth]{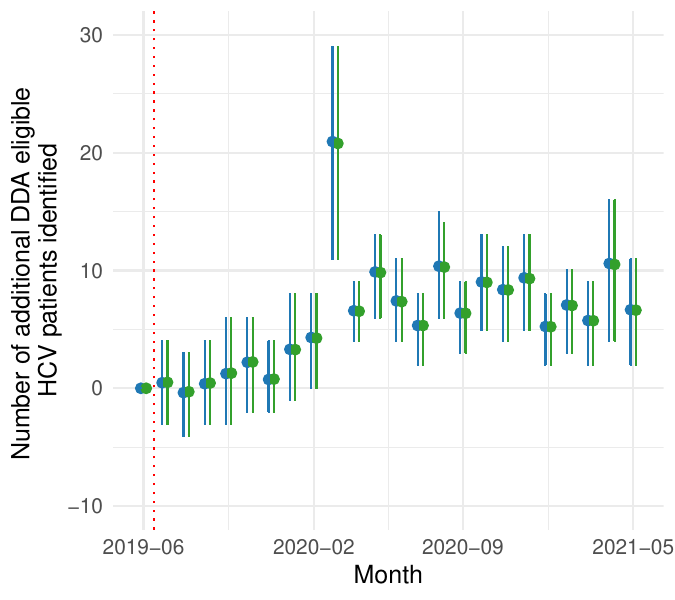} &
    \includegraphics[width=0.4\linewidth]{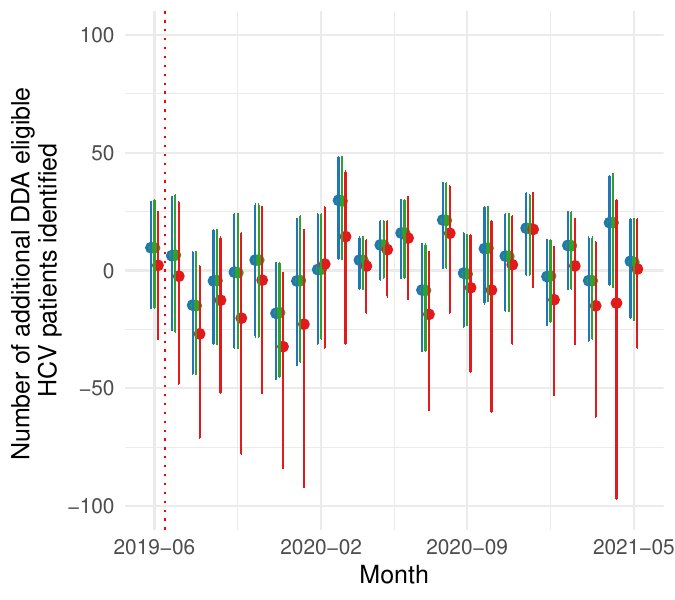} \\

    &\multicolumn{2}{c}{\includegraphics[width=0.7\linewidth]{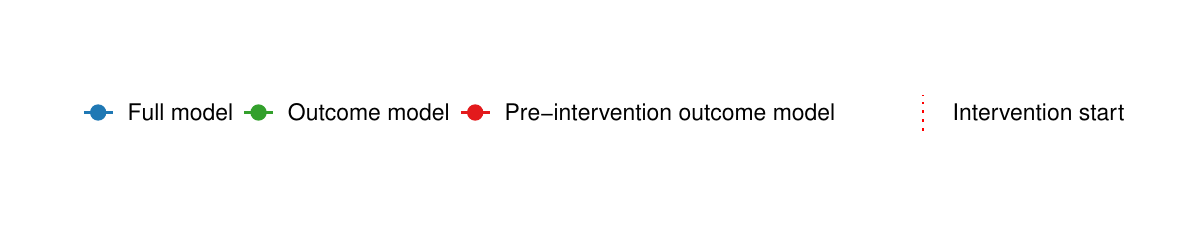}} \\
\end{tabular}
}
\caption{ Estimated individual intervention effects across time for the Operation Delivery Networks ‘Kent’, ‘Merseyside \& Cheshire’, and ‘South Yorkshire’. Abbreviations: Merseys. \& Che.: Merseyside \& Cheshire. $\rho$ is the assumed correlation between the potential outcomes using the Gaussian copula approach. }
\label{fig:ITErho}
\end{figure}

To investigate intervention effect heterogeneity, Figure \ref{fig:add_results}(a) displays the point estimates of $\tau_{it}$ (394 in total) for $\rho=0$.
To illustrate intervention effect patterns for single ODNs, Figure \ref{fig:ITErho} presents point estimates over time (along with 95\% CrIs) for three randomly selected ODNs, for $\rho=0$ and $\rho=1$. 
Although the estimated effects are generally positive, we observe that their magnitude varies considerably across time and ODNs. 
Some additional evidence of effect heterogeneity is presented in Figure D2 of Supplementary Material E, which shows the posterior densities of the dispersion parameters $\phi^0$ and $\phi^1$. 
Notably, we find that $\mathbb{P}(\phi_1 < \phi_0) = 0.831$, indicating greater variability around the mean under the intervention.


Next, we assess the extent to which effect heterogeneity is driven by cumulative exposure to the intervention and the presence of a lockdown. 
Figure \ref{fig:add_results}(b) presents $\exp\left(s\left(\sum_{j=1}^{t}a_{j}\right)\right)$, which represents the rate ratio $\omega_t \left(\bar{\boldsymbol{a}}_t\right)$ when there is no COVID-19 lockdown.
The estimated $\omega_t \left(\bar{\boldsymbol{a}}_t\right)$ are a concave function of the cumulative number of peer-months. 
There is little evidence of an increase in case-finding directly after intervention start. 
After around 9 peer-months we find evidence of a positive intervention effect, which peaks at approximately 16 peer-months and declines thereafter. 
This is in line with the findings of a previous United Kingdom-based qualitative study, which indicated that peers need time to integrate into the wider treatment program and the community before they can contribute \cite{bonnington_tensions_2017}. 
A potential explanation for the decline of the effect as cumulative exposure increases from moderate to high levels is the diminishing number of undiagnosed HCV-infected individuals, who are likely harder to reach. 

There is strong evidence that the intervention effect was amplified during the COVID-19 lockdown. 
This is evident from Figure \ref{fig:add_results}(a) and Figure \ref{fig:add_results}(c) showing that the magnitude of ITEs is considerably higher during the lockdown period and the showing that the posterior distribution of $\exp(\theta_1)$ is almost entirely above 1. 
Further evidence is provided in Figure \ref{fig:add_results}(d) that shows the posterior distribution of $\tau_c/\tau$ for $\rho=1$ and $\rho=0$. 
Although only $\tilde{N}_c=39$ of the total 394 ITEs correspond to the lockdown period, these ITEs account for 20.9\% (95\% CrI [12.9\%, 37.6\%]) of the total cumulative effect when $\rho=1$ (similar for $\rho=0$). 
This likely reflects the expanded role given to peers, for example in the delivery of DAA treatments, during the first national COVID-19 lockdown when the health system was under increased stress.

Lastly, we compare the results obtained from our approach with those from the approaches using only outcome data and using only pre-intervention outcome data.
Some comparisons are shown in Figures \ref{fig:Results_rho} (for $\tau$) and \ref{fig:ITErho} (for $\tau_{it}$, for three randomly selected ODNs). 
In Supplementary Material D, we further compare the point estimates of all ITEs and width of associated 95\%-CrIs across the three approaches. 
The results from the model using only outcome data (shown in green) closely resemble those from our method. 
This similarity is expected, given that relatively few peers were hired during the study period. 
As a consequence, the observed values of $A_{it}$ contribute limited information about the parameters shared ($\kappa_i$ and $\boldsymbol{\lambda}_i$) between the intervention assignment and potential outcomes models. 
Moreover, the alignment between the two approaches is reassuring, as large discrepancies in point estimates would suggest a conflict regarding the shared parameters between the two sources of information. 
In contrast, the results obtained from the pre-intervention outcome model (shown in red) differ substantially. 
Many of the point estimates of the ITEs differ from those obtained using our approach, and the 95\% CrIs are substantially wider than ours for the majority of ITEs. 
This is a major advantage of our approach compared to counterfactual prediction models: by incorporating post-intervention outcome data into the estimation process, parameters that are shared between models \eqref{eq:functional2} and \eqref{eq:functional3} are estimated with higher precision. 
We expect similar gains in precision in datasets that, like ours, involve units with short pre-intervention periods.

\section{Discussion}
\label{sec:Discussion3}

Motivated by a substantive application in the field of HCV, we proposed a novel Bayesian causal factor model for evaluating non-binary interventions with staggered adoption involving count data. 
Compared to state-of-the-art counterfactual imputation approaches, our method can improve credible interval coverage and efficiency by allowing post-intervention outcomes and intervention assignment to inform estimation of the factor and loadings parameters.
To our knowledge, our method is one of the few that take a fully Bayesian approach by modelling the intervention assignment mechanism. 
It further enables, using a copula approach, an easy means of assessing the sensitivity of estimates of ITEs (and related parameters) to assumptions regarding the joint distribution of potential outcomes. 
This is not possible when using counterfactual imputation models. 
Finally, our method provides uncertainty quantification for all causal parameters of interest.

We used our proposed approach to evaluate the effect of introducing peer supporters on the case-finding of HCV-infected individuals in England.
Considering estimates of both ITEs and ATEs (specifically rate ratios), we found strong evidence that peers increased case-finding.
Although we acknowledge limitations inherent to both types of estimands in the setting we are considering, we believe that, combined, they highlight the importance of this intervention in achieving HCV elimination targets.
Our analysis further suggested that intervention intensity was a driver of the intervention effect magnitude, where intensity refers to the cumulative number of peer-months of exposure up to that time. 
This finding demonstrates the importance of using information on treatment intensity when evaluating interventions, rather than treating them as binary. 
Finally, the effect of the peer intervention appeared to be especially strong during the first COVID-19 lockdown. 
To the best of our knowledge, this is a novel finding that has not been described in the literature before.
We believe this result demonstrates the potential of peer support to enhance engagement of HCV-infected individuals in treatment during periods when health services are under increased stress, and suggests that similar interventions may prove valuable in comparable situations in the future.

There are several interesting ideas for future research. 
One possibility is to consider more flexible copula models for the joint distribution of potential outcomes, such as vine copulas \cite{joe_tail_2010} or factor copulas \cite{krupskii_factor_2013}.
It would be interesting to investigate whether these more flexible copula models lead to estimates of ITEs with substantially wider credible intervals than our current approach. 
We have noted that the interpretation of ATEs and CATEs is challenging in studies such as ours, where the sample of units is exhaustive of the population. 
However, it might be possible to use these estimates to inform future interventions in different populations, such as different countries. 
The property of causal effects to be validly applied in a population external to the study population is known as transportability \cite{degtiar_review_2023}. 
We expect transportability to be particularly challenging in our setting due to the presence of temporal trends and the potential that the criteria for dividing a population into units (e.g., administrative regions) may not be comparable across populations.

There are also open questions regarding the effectiveness of the peers intervention.
First, it is possible to assess, using our method, the impact of peer support on alternative outcomes such as the number of HCV-infected individuals starting treatment and the number of HCV-infected individuals clearing the virus.
One challenge here is that recording of these outcomes is not mandatory, and thus information is more incomplete.
Another challenge relates to the fact that these outcomes are nested; i.e., HCV-infected individuals who start treatment are a subset of those identified as treatment eligible, and HCV-infected individuals who clear the virus are a subset of those starting treatment.
Applying our method to the number of individuals starting DAA treatment, for instance, is likely to suggest a positive intervention effect, since, as shown earlier, more DAA-eligible individuals are identified.
However, this does not necessarily imply that peer support increases the likelihood of treatment completion conditional on starting treatment.
A potential solution is to estimate separable direct effects \cite{samartsidis_bayesian_2024} instead of total effects.
Second, we are interested in comparing the demographics of the additional cases found thanks to peers to the demographics of the remaining cases.
This is useful for understanding which subpopulations benefit most from the intervention, as these models are primarily designed to address inequities in health system access.
We will investigate both questions in our future work.


\section*{Supplementary Material}
The Supplementary Material includes some further results for Section \ref{sec:BMCM}, the cross-validation algorithm, a description of a simulation study assessing the effect of the cut posterior approach in a small sample setting, and supplementary results of the real data application.

\newpage
\bibliographystyle{unsrtnat} 
\bibliography{references}        

\newpage

\end{document}